\documentclass{article}

\usepackage[numbers]{natbib}
\usepackage{rotating}
\usepackage{graphicx}
\def\be{\begin{equation}}
\def\ee#1{\label{#1}\end{equation}}

\def\lb{\label}
\newcommand{\ben}{\begin{eqnarray}}
\newcommand{\een}{\end{eqnarray}}
\def\bx{\mathbf{x}}

\def\bv{\mathbf{v}}
\def\bp{\mathbf{p}}
\def\bk{\mathbf{k}}
\def\k{\textsf{k} }
\def\br{\mathbf{r}}

\begin{document}

\title{Analysis of Jeans Instability from the Boltzmann Equation\footnote{This paper is dedicated to Professor Ingo M\"uller on the occasion of his eightieth  birthday.}}

\author{Gilberto M. Kremer\footnote{kremer@fisica.ufpr.br}
\\Departamento de F\'{\i}sica, Universidade Federal do Paran\'a\\ Caixa Postal 19044, 81531-980 Curitiba, Brazil}

\maketitle

\begin{abstract}
The dynamics of self-gravitating fluids is analyzed within the framework of a collisionless Boltzmann equation in the presence of gravitational fields and Poisson equation. Two cases are analyzed: a system with baryonic and dark matter in a static universe and a single system in an expanding universe. The amplitudes of the perturbed distribution functions are considered as a linear combination  of the collision invariants of the Boltzmann equation. For the system of baryonic and dark matter, the Jeans mass of the combined system is smaller than the one of the single system indicating that a smaller mass is needed to initiate the collapse. For the single system in an expanding universe it is not necessary to make use of Jeans "swindle"and it shown that for small wavelengths the density contrast oscillates while for large wavelengths it grows with time and the Jeans instability emerges.
\end{abstract}
 \section{Introduction}

 The search for structure formation from gas clouds is an old subject that goes back to 1902 when Jeans \cite{b1} used the system of phenomenological equations of mass and momentum densities together with the Poisson equation and showed that small perturbations in the mass density, pressure, velocity and gravitational potential in a static background with wavenumber smaller than the Jeans wavenumber  could evolve with time.   In terms of balance of forces  the fluctuations can grow in time if  the inwards directed gravitational force is larger than the outwards directed internal pressure of the gas.

 Jeans theory describes the gravitational instability of self-gravitating systems by searching for conditions that small perturbations can grow and leads to a collapse of the system \cite{b2,b3,b4,b5,b6,b8,b9,b10}.
 It was formulated before the knowledge of the Universe expansion and one has to take into account the Jeans ``swindle", which imposes that the Poisson equation is valid only for the perturbations,  since the background solution of constant mass density, pressure, gravitational potential and vanishing velocity satisfy the balance equations of mass and momentum densities, but not the Poisson equation.

The analysis of the small perturbations by using the same phenomenological equations of Jeans  but taking into account the expansion of the Universe was due to Bonnor \cite{b7} in 1957. The background solutions in this case satisfy the balance equations of mass and momentum densities and Poisson equation. In this context there is no necessity to invoke Jeans ``swindle" (see e.g. \cite{b2,b4,b7}).

Another way to examine Jeans instability is to consider the system of equations composed by the Newtonian version of the Boltzmann  equation and Poisson equation (see e.g. \cite{b4,b5,b11,cap,RG,GR}). In these works it is considered a static Universe and Jeans ``swindle". The dispersion relation follows by taking into account small perturbations of the equilibrium values of the distribution function and  gravitational potential.

Today it is well known that the matter content of the Universe is composed by baryonic matter which comprises atoms of all categories and dark matter which does not emit or interact  with electromagnetic radiation. The ratio baryonic matter to dark matter estimated is about one to five.
Dark matter plays a key role in structure formation because it feels only the force of gravity: the gravitational Jeans instability which allows compact structures to form is not opposed by any force, such as radiation pressure. As a result, dark matter begins to collapse into a complex network of dark matter halos well before ordinary matter, which is impeded by pressure forces. Without dark matter, the epoch of galaxy formation would occur substantially later in the universe than is observed.

The aim of this work is to  obtain the collapse criterion from the Newtonian version of the Boltzmann  equation and Poisson equation for a system of baryonic matter and dark matter and for a single system in an expanding spatially flat Universe. In both cases the amplitudes of the perturbed distribution functions are taken as a linear combination  of the collision invariants of the Boltzmann equation. For the system of baryonic and dark matter it is shown that the mass needed to initiate the collapse is smaller than the one of a single constituent and that the ratio of the dispersion velocities have a significant role in the structure formation.  For the single system  the equilibrium distribution function is written in a comoving frame which takes into account Hubble's law. Furthermore, the mass density is a solution the Friedmann and acceleration equations for a pressureless fluid. There is no necessity to use Jeans ``swindle", since the equilibrium distribution function and gravitational potential satisfy both the Boltzmann and Poisson equations. For this case the Jeans instability is connected with the growth  of the density contrast for large wavelengths, which is a parameter related with the local increase of matter density.

\section{Basic equations}

In the phase space spanned by the space and velocity coordinates $(\br,\bv)$ a state of a non-relativistic gas is characterized by the one-particle distribution function $f\equiv f(\br,\bv,t)$, such that $f(\br,\bv,t)d^3r d^3v$ gives the number of the particles in the volume element $d^3r$ about the position $\br$ and with velocity in the range $d^3v$ about $\bv$. The Boltzmann equation governs the space-time evolution of the one-particle distribution function $f$  in the phase space. In the presence of a gravitational potential $\Phi$ and in the absence of collisions between the particles, the Boltzmann equation is given by (see e.g. \cite{GMK,CC})
\ben\lb{1}
\partial_t f+\bv\cdot\nabla f -\nabla\Phi\cdot\partial_\bv f=0.
\een
Here the force per unit mass which acts on a particle is only of gravitational nature ${\bf F}=-\nabla\Phi$.

The Boltzmann equation (\ref{1}) follows also from a relativistic version of it in the presence of gravitational fields, where a one-particle distribution function $f\equiv f(\br,\bp,t)$  in the phase space spanned by the space coordinates and momenta $(\br,\bp)$ satisfies the Boltzmann equation (see e.g. \cite{CK})
\ben\lb{2}
p^{\mu}\partial_\mu f-\Gamma_{\mu\nu}^{i}p^{\mu}p^{\nu}\partial_{p^i} f=0.
\een
Above the mass-shell condition $p^\mu p_{\mu}=m^2c^2$ -- with $m$ denoting the particle rest mass -- was considered. In the non-relativistic  Newtonian limiting case $p^0 \rightarrow mc$, $\bp\rightarrow m\bv$ and the Christoffel symbol $\Gamma_{00}^i\rightarrow\nabla^i\Phi/c^2$ so that (\ref{2}) reduces to (\ref{1}).

The connection of the gravitational potential with the one-particle distribution function is given by Poisson equation
\ben\lb{3}
\nabla^2\Phi=4\pi G\rho=4\pi G\int mfd^3v,
\een
where $G$ denotes the gravitational constant and $\rho$ is the mass density of the fluid.

The analysis of cosmological problems is based on the cosmological principle which states that the Universe at large scales is spatially homogeneous and isotropic. The metric which satisfies this principle is the Friedmann-Lama\^itre-Robertson-Walker metric. For a spatially flat Universe the line element in this metric is written as
$ds^2=(cdt)^2-a(t)^2\left(dx^2+dy^2+dz^2\right),$ where $a(t)$ is the cosmic scale factor.

Einstein's field equations for a Universe dominated by a perfect fluid reduce to two coupled differential equations (see e.g. \cite{b2,b3,b4})
\ben\lb{5a}
\left(\frac{\dot a}{a}\right)^2=\frac{8\pi G}{3}\rho,\qquad \frac{\ddot a}{a}=-\frac{4\pi G}{3}\left(\rho+\frac{p}{c^2}\right),
\een
which are known as the Friedmann and acceleration equations, respectively. Above the dot denotes the derivative with respect to time and $\rho,p$ are the mass density and the pressure of the sources of the gravitational field, respectively.

In this work we are interested in matter  dominated Universe, where the sources of the gravitational field are pressureless $p=0$. In this case equations (\ref{5a}) can be solved for the mass density as a function of the cosmic scale factor, yielding
\ben\lb{6}
\rho=\rho_0\left(\frac{a_0}a\right)^3,
\een
where $\rho_0$ and $a_0$ are the values of the mass density and cosmic scale factor at $t=0$, respectively.

Another relationship which will be used below is Hubble's law  $\dot\br=(\dot a/a)\br$ which relates the recession velocity of an object  $\dot\br$ with its physical distance $\br$ in an expanding Universe. The quantity $H=\dot a/a$ is known as Hubble parameter.

\section{Jeans instability for systems of one component}

We start by analyzing the Jeans instability for systems with one component. This subject  was discussed in some books, see e.g. \cite{b4,b5}, but here a new methodology is introduced, which is based on the collision invariants of the Boltzmann equation \cite{CC}. The collisionless Boltzmann (\ref{1}) and Poisson (\ref{3}) equations imply a coupled system of equations for the distribution function $f(\br, \bv, t)$ and gravitational potential $\Phi(\br, t)$, which we reproduce below:
\ben\lb{4}
\partial_t f+\mathbf{v}\cdot\nabla f-\nabla \Phi\cdot\partial_\mathbf{v} f=0,
\qquad
\nabla^2\Phi=4\pi G\int mfd^3v.
\een

At equilibrium the distribution function and the gravitational potential of a self-gravitating system do not depend on time and the collisionless Boltzmann equation (\ref{4})$_1$ is satisfied if the distribution function depends only  of  the velocity of the particles $f_0(\bv)$ and the gravitational potential gradient vanishes $\nabla\Phi_0=0$. Note that the condition $\nabla\Phi_0=0$ may follow  from symmetry considerations, because in a homogeneous  system  there is no preference in the direction of the gravitational potential gradient. The  same restriction for the gravitational potential gradient  is also considered for a system described by the balance equations of mass density and fluid velocity, since constant values of pressure and density and vanishing fluid velocity and gravitational potential gradient satisfy these equations. However, the condition $\nabla\Phi_0=0$ does not satisfy the Poisson equation (\ref{4})$_2$ due to the fact that its right-hand side refers to the mass density of the self-gravitating system. In order to remove this inconsistency one makes use of the Jeans ``swindle", which considers that the Poisson equation is valid only for the perturbed distribution function and perturbed gravitational potential.

We require that the equilibrium state is subjected to small perturbations $h(\br, \bv, t)$ and $\Phi_1(\br,t)$:
\ben\lb{5}
f(\br, \bv, t)=f_0(\bv)\left[1+h(\br, \bv, t)\right],\qquad \Phi(\br, t)=\Phi_0+\Phi_1(\br,t).
\een
The equilibrium value of the gravitational potential is considered as a constant in order to fulfill the condition $\nabla\Phi_0=0$ and the equilibrium distribution function is the Maxwellian one
\ben\lb{6a}
f_0(\mathbf{v})=\frac{\rho}m\frac{e^{-{v^2}/{2\sigma^2}}}{(2\pi\sigma^2)^{3/2}},
\een
where $\sigma=\sqrt{kT/m}$ is the  dispersion (thermal) velocity and $k$, $T$ denote the Boltzmann constant and the temperature of the system, respectively. Note that the collisionless Boltzmann equation (\ref{4})$_1$ is a differential equation for the distribution function and its solution depends on the problem we are analyzing. Here we are interested in analyzing disturbances which can be represented by plane waves so that we have required that  the distribution function is a sum of an equilibrium distribution function plus an arbitrary function which characterizes a plane wave solution of small amplitude.

We insert the representations (\ref{5}) into  the Boltzmann  and Poisson  equations (\ref{4}) and get the following system of equations for the perturbed quantities $h$ and $\Phi_1$:
\ben\lb{7}
f_0\left[\partial_t h+\mathbf{v}\cdot\nabla h\right]
-\nabla \Phi_1\cdot\partial_\mathbf{v} f_0=0,
\qquad
\nabla^2\Phi_1=4\pi G\int m f_0hd^3v.
\een
In (\ref{7})$_1$ we have neglected  products of $\nabla \Phi_1$ with $h$ and $\partial_\mathbf{v}h$, since $h$ and $\Phi_1$ are supposed as small quantities. Furthermore, we have considered that  the Poisson equation refers only to the perturbed values of the distribution function and gravitational potential and taken $\nabla\Phi_0=0$.

The perturbations $h$ and $\Phi_1$ are now represented by plane waves of frequency $\omega$ and wavenumber vector $\bk$ such that
\ben\lb{8}
h(\mathbf{r},\mathbf{v},t)=h_1(\mathbf{v})\exp\left[i\left(\bk\cdot\mathbf{r}-\omega t\right)\right],\qquad
\Phi_1(\mathbf{r},t)=\phi\exp\left[i\left(\bk\cdot\mathbf{r}-\omega t\right)\right].
\een
 In the above equations $\phi$ is a constant amplitude while $h_1(\mathbf{v})$ is given as a combination of the collision invariants of the Boltzmann equation, namely: 1, $\mathbf{v}$ and $\mathbf{v}^2$ (see e.g. \cite{CC,GMK}). Its expression is written as
\ben\lb{9}
h_1(\mathbf{v})=A+\mathbf{B}\cdot\mathbf{v}+D\mathbf{v}^2,
\een
where $A$, $\bf B$ and $D$ are unknown constants.

The insertion of (\ref{8}) and (\ref{9})  together with (\ref{6a}) into the perturbed Boltzmann  and Poisson equations (\ref{7}) lead to
\ben\lb{10}
\left(A+\mathbf{B}\cdot\mathbf{v}+D\mathbf{v}^2\right)\left[\omega -\bk\cdot\mathbf{v}\right]-\bk\cdot\mathbf{v}\frac{\phi}{\sigma^2}=0,
\qquad \k^2\phi+4\pi G\left(A+3\sigma^2D\right)\rho=0,
\een
where $\k=\vert\bk\vert$.

Now a system of equations for $A$, $B=\bf B\cdot k$, $D$ and $\phi$ can be obtained from the multiplication of (\ref{10})$_1$ by the collision invariants $\left( 1, \mathbf{v}, \mathbf{v}^2\right)$ and integration of the resulting equations, yielding
\ben\lb{11}
\omega \left(A+3\sigma^2D\right)-\sigma^2B=0,
\qquad
\omega B-\left[A+5\sigma^2D+\frac\phi{\sigma^2}\right]\k^2=0,\qquad
\omega \left(3A+15\sigma^2D\right)-5\sigma^2B=0.
\een

The system of equations  (\ref{11}) together with (\ref{10})$_2$ has a non-trivial solution if the determinant of the coefficients  $A$, $B$, $D$ and $\phi$ vanishes, which implies the dispersion relation:
\ben\lb{12}
\omega_*\left(1-\k_*^2+\omega_*^2\right)=0.
\een
Above the dimensionless wavenumber modulus $\k_*$ and frequency $\omega_*$ are given by the relationships
\ben\lb{13}
\k_*=\frac{v_s \k}{\sqrt{4\pi G\rho}}=\frac{\k}{\k_J},\qquad \omega_*=\frac{\omega}{\sqrt{4\pi G\rho}}.
\een
Here we have introduced  the adiabatic sound speed $v_s=\sqrt{5/3}\,\sigma$ and  the Jeans wavenumber $\k_J=\sqrt{4\pi G\rho}/v_s$.

We have the following solutions for the dimensionless frequency from the dispersion relation  (\ref{12}):  $\omega_*=0$ and
\ben\lb{14}
\omega_*=\pm\sqrt{\frac{\k^2}{\k_J^2}-1}.
\een
From the above equation we may infer that for $\k>\k_J$ (small wavelengths), $\omega_*$ is a real quantity and the perturbations -- due to the factor $\exp\left(-i\omega t\right)$ -- will propagate as harmonic waves in time. However, if  $\k<\k_J$ (big wavelengths), $\omega_*$ is a pure imaginary quantity and one of the perturbations will grow with time, while the other will decay with time. The perturbation which grows with time is associated with Jeans instability.

A very simple model is described in the literature to understand the Jeans instability. Consider a mass $M$ enclosed in a spherical volume of radius $\lambda$ in which there exists a mass density inhomogeneity. This inhomogeneity will grow if the gravity force $F_G$ per unit mass is greater than the opposed pressure force $F_P$ per unit of mass, i.e.,
\ben\lb{15}
F_G= \frac{GM}{\lambda^2}\propto \frac{G\rho\lambda^3}{\lambda^2}=G\rho\lambda>F_P\propto \frac{p\lambda^2}{\rho\lambda^3}\propto \frac{v_s^2}\lambda,\qquad\hbox{since} \qquad v_s^2\propto \frac{p}\rho.
\een
Now identifying  the Jeans wavelength in terms of the Jeans wavenumber $\lambda_J=2\pi/\k_J={2\pi v_s}/{\sqrt{4\pi G\rho}}$ and $\lambda=2\pi/\k$,  the inequality $\k<\k_J$ follows.

\section{Jeans instability for systems with two components}

In this section we shall analyze Jeans instability by taking into account two collisionless Boltzmann equations, one for the baryonic matter and another for the dark matter which are connected with the  Poisson equation. We shall use the indices $b$ and $d$ for the  baryonic and dark matter, respectively. The corresponding evolution equations for the distribution functions of baryonic matter $f_b\equiv f(\bx,\bv_b,t)$ and dark matter $f_d\equiv f(\bx,\bv_d,t)$ are given by the collisionless Boltzmann equations
\ben\lb{16}
\partial_t f_b+\bv_b\cdot\nabla f_b -\nabla\Phi\cdot\partial_{\bv_b} f_b=0,\qquad
\partial_t f_d+\bv_d\cdot\nabla f_d -\nabla\Phi\cdot\partial_{\bv_d} f_d=0,
\een
 which are connected with the Poisson equation
\ben\lb{17}
\nabla^2\Phi=4\pi G \left(\int m_b f_b d^3v_b+\int m_d f_d d^3v_d\right)
=4\pi G(\rho_b+\rho_d).
\een
Above $(m_b,m_d)$ and $(\rho_b,\rho_d)$ are the masses and mass densities of the baryonic and dark matter, respectively.

Following the same methodology of the previous section we suppose that the distribution functions $ f(\bx,\bv_b,t)$, $f(\bx,\bv_d,t)$ and the potential $\Phi$ are subjected to small perturbations $h_b(\mathbf{r},\mathbf{v}_b,t)$, $h_d(\mathbf{r},\mathbf{v}_d,t)$ and $\Phi_1(\mathbf{r},t)$ from their equilibrium values $f_b^0({\bf v}_b)$, $f_d^0({\bf v}_d)$ and $\Phi_0$, respectively. Hence we write
\ben\lb{18}
f(\mathbf{r},\mathbf{v}_b,t)=f_b^0({\bf v}_b)\left[1+h_b(\mathbf{r},\mathbf{v}_b,t)\right],
\qquad
f(\mathbf{r},\mathbf{v}_d,t)=f_d^0({\bf v}_d)\left[1+h_d(\mathbf{r},\mathbf{v}_d,t)\right],
\qquad
\Phi(\mathbf{r},t)=\Phi_0+\Phi_1(\mathbf{r},t).
\een
where the equilibrium distribution functions  are the Maxwellians
\ben\lb{19}
f_b^0(\mathbf{r},\mathbf{v}_b,t)=\frac{\rho_b}{m_b}\frac{e^{-\mathbf{v}_b^2/2\sigma_b^2}}{(2\pi\sigma_b^2)^{3/2}},
\qquad
f_d^0(\mathbf{r},\mathbf{v}_d,t)=\frac{\rho_d}{m_d}\frac{e^{-\mathbf{v}_d^2/2\sigma_d^2}}{(2\pi\sigma_d^2)^{3/2}}.
\een
Here $\sigma_b=\sqrt{kT_b/m_b}$ and $\sigma_d=\sqrt{kT_d/m_d}$ are the dispersion velocities of the baryonic and dark matter which are associated with their temperatures $T_b$ and $T_d$, respectively.

From the insertion of the representations (\ref{18}) into the Boltzmann (\ref{16}) and Poisson (\ref{17}) equations yields
\ben\lb{20a}
&&f_b^0\left[\partial_t h_b+\mathbf{v}_b\cdot\nabla h_b\right]
-\nabla \Phi_1\cdot\partial_{\mathbf{v}_b} f_b^0=0,\qquad
f_d^0\left[\partial_t h_d+\mathbf{v}_d\cdot\nabla h_d\right]
-\nabla \Phi_1\cdot\partial_{\mathbf{v}_d} f_d^0=0,
\\\lb{20b}
&&\nabla^2\Phi_1=4\pi G\left(\int m_b f_b^0h_bd^3v_b+\int m_d f_d^0h_dd^3v_d\right).
\een
Here we followed the same methodology  which was used to obtain the system of equations (\ref{7}) and considered the Jeans "swindle".

We represent the perturbations as plane waves of frequency $\omega$ and wavenumber vector ${\bf k}$
\ben\lb{21}
h_b(\mathbf{r},\mathbf{v}_b,t)=h_b^1(\mathbf{v}_b)
\exp\left(i\mathbf{k}\cdot\mathbf{r}-i\omega t\right),
\quad
h_d(\mathbf{r},\mathbf{v}_d,t)=h_d^1(\mathbf{v}_d)
\exp\left(i\mathbf{k}\cdot\mathbf{r}-i\omega t\right),\quad
\Phi_1(\mathbf{r},t)=\phi\exp\left(i\mathbf{k}\cdot\mathbf{r}-i\omega t\right),
\een
where the amplitude $\phi$ is constant, while $h_b^1$ and $h_d^1$ are given in terms of the collision invariants of the Boltzmann equations $\left( 1, \mathbf{v}_b, \mathbf{v}_b^2\right)$ and $\left( 1, \mathbf{v}_d, \mathbf{v}_d^2\right)$:
\ben\lb{22}
h_b^1({\bf v}_b)=A_b+\mathbf{B}_b\cdot\mathbf{v}_b+D_b\mathbf{v}_b^2,\qquad h_d^1({\bf v}_d)=A_d+\mathbf{B}_d\cdot\mathbf{v}_d+D_d\mathbf{v}_d^2.
\een
Above $A_b,A_d, \mathbf{B}_b,\mathbf{B}_d, D_b, D_d$ are constants

Now the insertion of (\ref{21}) together with (\ref{22}) and (\ref{19})  into (\ref{20a}) and  (\ref{20b}) leads to
\ben\lb{23a}
&&\left(A_\alpha+\mathbf{B}_\alpha\cdot\mathbf{v}_\alpha+D_\alpha\mathbf{v}_\alpha^2\right)\left[\omega -\mathbf{k}\cdot\mathbf{v}_\alpha\right]
-\mathbf{k}\cdot\mathbf{v}_\alpha\frac{\phi}{\sigma_\alpha^2}=0,\qquad \alpha=b,d
\\\lb{23b}
&&\k^2\phi+4\pi G\left[\left(A_b+3\sigma_b^2D_b\right)\rho_b+\left(A_d+3\sigma_d^2D_d\right)\rho_d\right]=0.
\een

In order to get a system of algebraic equations for  $A_b,A_d, B_b=\mathbf{B}_b\cdot \mathbf{k},B_d=\mathbf{B}_d\cdot \mathbf{k}, D_b, D_d$ and  $\phi$ we multiply (\ref{23a})  by the collision invariants $\left( 1, \mathbf{v}_b, \mathbf{v}_b^2\right)$ and $\left( 1, \mathbf{v}_d, \mathbf{v}_d^2\right)$, respectively, and integrate the resulting equations, yielding for $(\alpha=b,d)$
\ben\lb{24a}
\omega \left(A_\alpha+3\sigma_\alpha^2D_\alpha\right)-B_\alpha\sigma_\alpha^2=0,\qquad
\omega B_\alpha
-\left[A_\alpha+5\sigma_\alpha^2D_\alpha
+\frac{\phi}{\sigma_\alpha^2}\right]\k^2=0,\qquad
\omega \left(3A_\alpha+15\sigma_\alpha^2D_\alpha\right)-5B_\alpha\sigma_\alpha^2=0.
\een

A non-trivial solution of the system of equations (\ref{23b}) and (\ref{24a}) is found  if the determinant of the coefficients $A_b,A_d,B_b,B_d,D_b,D_d$, $\phi$ vanishes, which implies the dispersion relation:
\ben\lb{25}
\omega_*^4+\left[1+\frac{\rho_b}{\rho_d}-\left(1+\frac{\sigma_b^2}{\sigma_d^2}\right)\k_*^2\right]
\omega_*^2+\frac{\sigma_b^2}{\sigma_d^2}\left[\k_*^2-1-\frac{\rho_b\sigma_d^2}{\rho_d\sigma_b^2}\right]
\k_*^2=0.
\een
Here we have introduced the dimensionless wavenumber $\k_*$ and frequency $\omega_*$ defined in terms of the dark matter Jeans wavenumber $\k_J^d=\sqrt{4\pi G\rho_d}/v_s^d$ -- with $v_s^d=\sqrt{5/3}\,\sigma_d$ denoting the dark matter sound speed  --  since as it was explained in the introduction the dark matter begins to collapse into a complex network of dark matter halos well before ordinary matter. The dimensionless wavenumber and frequency read
\ben\lb{26}
\k_\ast=\frac{\k}{\k_J^d}=\frac{\k \,v_s^d}{\sqrt{4\pi G\rho_d}},\qquad \omega_\ast=\frac{\omega}{\sqrt{4\pi G\rho_d}}.
\een

We note that the dispersion relation (\ref{25}) is a function of two ratios $\rho_d/\rho_b$ and $\sigma_d/\sigma_b$. The mass density ratio $\rho_d/\rho_b$ can be  associated with the present value of the density parameter ratio $\rho_d/\rho_b=\Omega_d/\Omega_b\approx5.5$ \cite{val}, since this ratio has not changed considerably  during the evolution of the universe. For the dispersion velocities ratio $\sigma_d/\sigma_b$ there is no fixed value. One value we shall use is taken from Ref. \cite{mw} where Milky Way-like galaxy simulations including both baryonic and dark matter were performed. From this work we have inferred that in one of the simulations where Maxwellian distributions are considered the ratio is given by $\sigma_d/\sigma_b=170/93\approx1.83$.

From the dispersion relation  (\ref{25}) we have the following solutions for the dimensionless frequency:
\ben\lb{27}
\omega_*&=&\pm\frac{\sigma_b}{\sqrt2\sigma_d}\sqrt{\k_*^2\left(1+\frac{\sigma_d^2}{\sigma_b^2}\right)
-\frac{\sigma_d^2}{\sigma_b^2}\left(1+\frac{\rho_b}{\rho_d}\right)
\pm\Delta},\\
\Delta&=&\sqrt{\left[\k_*^2\left(1+\frac{\sigma_d^2}{\sigma_b^2}\right)-\frac{\sigma_d^2}{\sigma_b^2}
\left(1+\frac{\rho_b}{\rho_d}\right)\right]^2-4\frac{\sigma_d^2}{\sigma_b^2}
\left[\k_*^4-\left(1+\frac{\rho_b\sigma_d^2}{\rho_d\sigma_b^2}\right)\k_*^2\right]}.
\een

Without the baryonic matter, i.e. for $\rho_b=\sigma_b=0$,  the dispersion relation (\ref{25})   reduces to:
\ben\lb{28}
\omega_*^4+\left[1-\k_*^2\right]\omega_*^2=0,
\een
which has the solutions $\omega_*=0$ and $\omega_*=\pm\sqrt{\k^2/\k_J^2-1}$, i. e., we recover the Jeans solution for one component.

Two solutions of (\ref{27}) provide  real values for $\omega_*$ for any $\k_*$ so that we  have harmonic waves in time for these two solutions. However the other two solutions furnish imaginary values for $\omega_*$ for some values of $\k_*$ so that instead of harmonic waves the amplitude of the disturbance will grow or decay. The one which grows is associated with Jeans instability. The value of $\k_*$ where $\omega_*$ changes from the imaginary value to the real value is obtained by taking $\omega_*$ equal to zero in (\ref{25}) and we get
\ben\lb{29}
\k_*=\frac{\k_J^{\rm db}}{\k_J^{\rm d}}=\sqrt{1+\frac{\rho_b\sigma_d^2}{\rho_d\sigma_b^2}}=\frac{\lambda_J^{\rm d}}{\lambda_J^{\rm db}}.
\een
This equation is interpreted as the ratio of two Jeans wavenumbers, the one $\k_J^{\rm db}$ refers to the system dark-baryonic matter while  $\k_J^{\rm d}$ to the dark matter.

We can also analyze the amount of mass which is necessary to initiate the collapse, which is the Jeans mass contained in a sphere of diameter equal to the wavelength $\lambda=2\pi/\k$. Hence, we can build the ratio of Jeans masses, one for the system dark-baryonic matter $M_J^{\rm db}$ and another for the dark matter alone $M_J^{\rm d}$, namely
\ben\lb{30}
\frac{M_J^{\rm db}}{M_J^{\rm d}}=\frac{\rho_b+\rho_d}{\rho_d}\left(\frac{\lambda_J^{\rm db}}{\lambda_J^{\rm d}}\right)^3=\left(1+\frac{\rho_b}{\rho_d}\right)
\left(\sqrt{1+\frac{\rho_b\sigma_d^2}{\rho_d\sigma_b^2}}\right)^{-3}.
\een

\begin{table}[h]
\label{tab:a}
\begin{tabular}{|c|c|c|c|c|c|c|c|}
\hline
$\sigma_d/\sigma_b$&1.00&1.20&1.40&1.60&1.83 &2.00 &2.20\\\hline
$M_J^{\rm db}/M_J^{\rm d}$&0.9199&0.8338&0.7481&0.6662&0.5791&0.5206&0.4585\\
\hline
\end{tabular}
\caption{Ratio of Jeans masses $M_J^{\rm db}/M_J^{\rm d}$ as function of the ratio of the dispersion velocities $\sigma_d/\sigma_b$ for $\rho_d/\rho_b=5.5$.}
\end{table}

In Table \ref{tab:a} the ratio of the Jeans masses of the system baryonic-dark matter and dark matter are given as functions of the dispersion velocities ratio for fixed values of the mass densities ratio. From this table we infer that if we  increase the dispersion velocities ratio, the mass needed to begin the collapse becomes smaller in comparison with the mass where only one constituent is present. This can be understood, since when the ratio  $\sigma_d/\sigma_b$ is large the dispersion velocity of the baryonic matter is smaller than the one of the dark matter and the baryonic matter hardly overcome the escape velocity of a given gravitational field.

\section{Jeans instability in an expanding Universe }

In this section we shall analyze Jeans instability by taking into account the collisionless Boltzmann and Poisson equations (\ref{7}) but in an expanding Universe where a pressureless fluid is the source of the gravitational field.

Now the equilibrium distribution function is written in a comoving frame as
\ben\lb{31}
f_0(\mathbf{r},\mathbf{v},t)=\frac{\rho}m\frac1{(2\pi\sigma^2)^{3/2}}\exp\left(-\frac{\left(\mathbf{v}-\frac{\dot a}{a}\mathbf{r}\right)^2}{2\sigma^2}\right),
\een
thanks to Hubble's law $\dot\br=(\dot a/a)\br$. Note that according to (\ref{6})
the mass density $\rho$  is only a function of time.

The gravitational potential function
\ben\lb{32}
\Phi_0(\mathbf{r},t)=\frac{2\pi}3G\rho r^2,
\een
and the equilibrium distribution function (\ref{31}) satisfy the Poisson and Boltzmann equations
\ben\lb{33}
\nabla^2\Phi_0=4\pi G\int mf_0d^3v=4\pi G\rho,\qquad \partial_t f_0+\mathbf{v}\cdot\nabla f_0-\nabla \Phi_0\cdot\partial_\mathbf{v} f_0=0,
\een
by taking into account the Friedmann and acceleration equations (\ref{5a}) and provided that the dispersion velocity is proportional to the inverse of the cosmic scale factor and does not depend on the spatial coordinates, namely $\sigma(t)\propto 1/a(t)$.

As in the previous sections we require that the  equilibrium state defined by the distribution function (\ref{31}) and gravitational potential (\ref{32}) is subjected to small perturbations characterized by $h(\mathbf{r},\mathbf{v},t)$ and $\Phi_1(\mathbf{r},t)$ such that
\ben\lb{35}
f(\mathbf{r},\mathbf{v},t)=f_0(\mathbf{r},\mathbf{v},t)\left[1+h(\mathbf{r},\mathbf{v},t)\right],
\qquad
\Phi(\mathbf{r},t)=\Phi_0(\mathbf{r},t)+\Phi_1(\mathbf{r},t).
\een
Note that (\ref{35}) is similar to  (\ref{5}), the  difference between them lies in the dependence of the equilibrium distribution function and gravitational potential at equilibrium, which now are functions of the space-time coordinates.

Now the insertion of (\ref{35}) into the collisionless Boltzmann  and Poisson equations (\ref{4}) leads the system of equations for $h$ and $\Phi_1$:
\ben\lb{36}
f_0\left[\partial_t h+\mathbf{v}\cdot\nabla h-\underline{\nabla \Phi_0\cdot\partial_\mathbf{v} h}\right]
-\nabla \Phi_1\cdot\partial_\mathbf{v} f_0=0,
\qquad
\nabla^2\Phi_1=4\pi G\int m f_0hd^3v.
\een
Note that the underlined term above does not show up in (\ref{7}), since it was supposed that $\nabla\Phi_0=0$. Above the products of $\nabla \Phi_1$ with $h$ and $\partial_\mathbf{v}h$ were also neglected.

The perturbations $h$ and $\Phi_1$ represented by plane waves of  wavenumber vector ${\bf q}/a(t)$, reads
\ben\lb{37}
h(\mathbf{r},\mathbf{v},t)=h_1(\mathbf{r},\mathbf{v},t)\exp\left(i\frac{\mathbf{q}\cdot\mathbf{r}}{a(t)}\right),\qquad
\Phi_1(\mathbf{r},t)=\phi(t)\exp\left(i\frac{\mathbf{q}\cdot\mathbf{r}}{a(t)}\right).
\een
The factor $1/a(t)$ in the wavenumber reflects the fact that the wavelength is stretched out in an expanding Universe. Here we cannot assume a harmonic wave in time, since the factors of the equations depend on time. Hence, $\phi(t)$ is an amplitude that depends on time, while $h_1(\mathbf{r},\mathbf{v},t)$ is given as a combination of the collision invariants of the Boltzmann equation in a comoving frame, i.e., 1, $\mathbf{v}-\frac{\dot a}{a}\mathbf{r}$ and $\left(\mathbf{v}-\frac{\dot a}{a}\mathbf{r}\right)^2$:
\ben\lb{38}
h_1(\mathbf{r},\mathbf{v},t)=A(t)+\mathbf{B}(t)\cdot\left(\mathbf{v}-\frac{\dot a}{a}\mathbf{r}\right)+D(t)\left(\mathbf{v}-\frac{\dot a}{a}\mathbf{r}\right)^2,
\een
where  $A(t)$, ${\bf B}(t)$ and $D(t)$ are also functions of time.

The methodology used here is the same as in the previous sections. We insert (\ref{37}) and (\ref{38}) together with (\ref{31}) and (\ref{32}) into the perturbed Boltzmann  and Poisson  equations (\ref{36}) and multiply the resulting equation from the Boltzmann equation by the collision invariants $\left( 1, \mathbf{v}-\frac{\dot a}{a}\mathbf{r}, \left(\mathbf{v}-\frac{\dot a}{a}\mathbf{r}\right)^2\right)$ and integrate. Hence we get the following  system of differential equations:
\ben\lb{39a}
&&\frac{dA(t)}{dt}+3\sigma^2\frac{dD(t)}{dt}+i\frac{\sigma^2}{a}B(t)-6\frac{\dot a}a\sigma^2D(t)=0,\qquad
\frac{dB(t)}{dt}+i\frac{q^2}a\left[A(t)+5\sigma^2D(t)+\frac{\phi(t)}{\sigma^2}\right]-\frac{\dot a}aB(t)=0,\qquad
\\\lb{39b}
&&3\frac{dA(t)}{dt}+15\sigma^2\frac{dD(t)}{dt}+i5\frac{\sigma^2}{a}B(t)-30\frac{\dot a}a\sigma^2D(t)=0,\qquad
\frac{q^2}{a^2}\phi(t)+4\pi G\left[A(t)+3\sigma^2D(t)\right]\rho=0.
\een
Above we have introduced $B(t)=\mathbf{B}(t)\cdot \mathbf{q}$.

In galaxy formation is usual to introduce the density contrast which is a parameter that indicates where there are local increase in the  matter density. It is defined by the ratio of perturbed and unperturbed mass densities  $\delta_\rho=\overline\rho/\rho$. Here the density contrast reads $\delta_\rho= A(t)+3\sigma^2D(t)$, and we can obtain from the system of differential equations (\ref{39a}) and (\ref{39b}) that $A$ is a constant (say $A=1$) and the following differential equation for the density contrast
\ben\lb{40}
 \tau^2\,\delta_\rho''+\frac4{3}\tau\,\delta_\rho'-\frac2{3}
\left(1-\frac{3\lambda_J^2}{5\lambda_0^2\tau^\frac23}\right)\delta_\rho-\frac{4\lambda_J^2}{25\lambda_0^2\tau^\frac23}=0,\qquad\hbox{where} \qquad\lambda_0=\frac{2\pi a_0}{q},\qquad \lambda_J=\frac{10\pi\sigma_0}{3\sqrt{4\pi G\rho_0}}.
\een
In the above equation we have used the relationship $a=a_0(6\pi G\rho_0 t^2)^{1/3}$ which follows from the integration of (\ref{5a}). Furthermore, $\tau=t\sqrt{6\pi G\rho_0}$ is a dimensionless time and the primes are derivatives with respect to $\tau$. The  solution of (\ref{40}) is given in terms of Bessel functions of first kind $J_{\pm\frac52}(\Lambda/\tau^\frac13)$ and reads
\ben\lb{41}
\delta_\rho=\tau^{-\frac16}\left[C_1J_{\frac52}\left(\frac{\Lambda}{\tau^\frac13}\right)+
C_2J_{-\frac52}\left(\frac{\Lambda}{\tau^\frac13}\right)\right]+\underline{\frac{2}{5}
\left(1+\frac{5\tau^\frac23}{3\Lambda^2}\right)},\qquad\hbox{where}\qquad \Lambda=\sqrt{\frac{18}{5}}\frac{\lambda_J}{\lambda_0}.
\een

\begin{figure}[h]
  \centerline{\includegraphics[width=350pt]{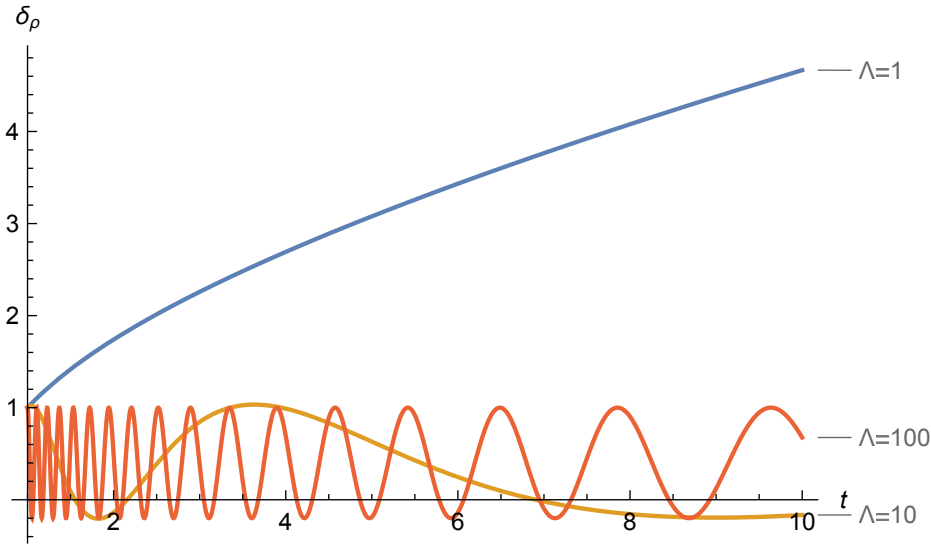}}\label{fig1}
  \caption{(color online) Density contrast $\delta_\rho$ as function of time $\tau$ for different values of the ratio $\Lambda=\sqrt{18/5}\lambda_J/\lambda_0$.}
\end{figure}

The solution (\ref{41}) differs from the one that comes from the analysis of the balance equations for mass, momentum and entropy densities in an expanding universe (see e.g. \cite{b4}) due to the underlined term which refers to the time evolution of the density contrast in a matter dominated universe. For large values of the ratio $\lambda_J/\lambda_0$ (small wavelengths) the Bessel functions imply oscillations of the density contrast while for small values of $\lambda_J/\lambda_0$ (large wavelengths) the first term in (\ref{41}) furnishes two solutions one decaying according to $1/\tau$ and another growing with $\tau^{2/3}$. However due to the underlined term with $\tau^{2/3}$ the density contrast will grow with time for small values of $\lambda_J/\lambda_0$ and this solution corresponds to Jeans instability. In Fig. \ref{fig1} it is shown  the time evolution of the density contrast for different values of $\Lambda$, i.e. for different values of the ratio $\lambda_J/\lambda_0$.

\section{Conclusions}

In this work we have analyzed Jeans instability using the collisionless Boltzmann and Poisson equations for a system of baryonic and dark matter in a static universe and a system of a single constituent in an expanding universe. For the system baryonic-dark matter the Jeans mass of the system is smaller than that of a single component indicating that the structures are formed earlier than in the latter case. Furthermore, the dispersion velocities ratio dark-matter/baryonic-matter has influence of the Jeans masses, since small dispersion velocity of the baryonic matter in comparison with the one of the dark matter implies that the baryonic matter hardly overcome the escape velocity of a given gravitational field. For the single component in  an expanding Universe the background solutions for the distribution function and gravitational potential satisfy both Boltzmann and Poisson equations without the introduction of Jeans ``swindle". The equilibrium distribution function is written in a comoving frame that takes into account Hubble's law and the mass density is a solution of Friedmann and acceleration equations for a pressureless fluid.  The Jeans instability is connected with a parameter that indicates  the local increase in the  matter density, the so-called density contrast. It is shown that for large wavelengths the density contrast grows with time.

\section*{Acknowledgments}
This research was supported by the Conselho Nacional de Desenvolvimento Cient\'{\i}fico e Tecnol\'ogico (CNPq), Brazil.


\nocite{*}
\bibliographystyle{aipnum-cp}

\end{document}